\documentclass[%
 reprint,
 amsmath,amssymb,
 aps,
]{revtex4-2}

\usepackage{graphicx}
\usepackage{dcolumn}
\usepackage{bm}
\usepackage{multirow}
\usepackage{xcolor}
\usepackage{hyperref}

\begin{document}

\preprint{APS/123-QED}

\title{Rapid Mass Parameter Estimation of \\ Binary Black Hole Coalescences Using Deep Learning}

\author{Alistair McLeod}
\email{alistairmcleodresearch@gmail.com,\\
 22252522@student.uwa.edu.au}
\author{Daniel Jacobs}
\author{Chayan Chatterjee}
\author{Linqing Wen}
  \email{linqing.wen@uwa.edu.au}
\author{Fiona Panther}
\affiliation{Department of Physics, OzGrav-UWA, The University of Western Australia, 35 Stirling Hwy, Crawley, Western Australia 6009, Australia}

\date{\today}

\begin{abstract}

Deep learning can be used to drastically decrease the processing time of parameter estimation for coalescing binaries of compact objects including black holes and neutron stars detected in gravitational waves (GWs). As a first step, we present two neural network models trained to rapidly estimate the posterior distributions of the chirp mass and mass ratio of a detected binary black hole system from the GW strain data of LIGO Hanford and Livingston Observatories. Using these parameters the component masses can be predicted, which has implications for the prediction of the likelihood that a merger contains a neutron star. The results are compared to the `gold standard' of parameter estimation of gravitational waves used by the LIGO-Virgo Collaboration (LVC), LALInference. Our models predict posterior distributions consistent with that from LALInference while using orders of magnitude less processing time once the models are trained. The median predictions are within the 90\% credible intervals of LALInference for all predicted parameters when tested on real binary black hole events detected during the LVC's first and second observing runs. We argue that deep learning has strong potential for low-latency high-accuracy parameter estimation suitable for real-time GW search pipelines.

\end{abstract}

\maketitle

\section{\label{sec:level1}Introduction}

Gravitational waves are ripples in spacetime, first predicted by \citet{Einstein}, and first successfully detected by LIGO \cite{gw1} in 2015. The majority of events detected by the LIGO-Virgo Collaboration (LVC) \cite{ALIGO,AVIRGO} are binary black hole (BBH) mergers \cite{gwtc1, gwtc2, gwtc3}, but binary neutron star (BNS) mergers \cite{170817} and neutron star - black hole mergers (NSBH) \cite{nsbh} have also been detected. Most significant gravitational wave events were detected first in online real-time searches which use either matched filtering or burst search methods for signal filtering \cite{chichi, gstlal,pycbclive,mbta,cwb}. Detecting these events in real-time is of importance as it enables rapid follow-up observations of the event by electromagnetic telescopes. To this end, rapid parameter estimation is needed to assess the likelihood that the merger contains a neutron star that is considered more likely to be associated with electromagnetic radiation.

These compact binary systems are described by a 15 dimensional parameter space, consisting of intrinsic attributes of the two merging compact objects (component mass, spin etc), and extrinsic parameters such as the orientation of the merging system to a detector. Once a gravitational wave is detected in detector data by a real-time search pipeline, the pipeline performs parameter estimation on the detected event \cite{chichi}. Parameter estimation is currently performed as a Bayesian analysis with Markov Chain Monte Carlo and nested sampling of the parameter posteriors \cite{bayes_intro, LALInference, bilby, rift}. While this is extremely accurate, it is also extremely computationally intensive, and can take hours or days for a full parameter estimation of a binary black hole system to be calculated \cite{BerryChristopherP.L2015PEFB}. This is insufficient for BNS mergers, as prompt electromagnetic radiation could be emitted within seconds of the merger \cite{grb}. Rapid parameter estimation of gravitational wave signals is needed to help capture these prompt electromagnetic emissions.

Several machine learning systems have been developed for BBH parameter estimation \cite{cuoco, chatterjee_localise, haegel}. Neural networks of varying architectures have been shown to be able to accurately approximate Bayesian posteriors for source parameters as a Gaussian mixture \cite{ChuaAlvinJ}, and even perform full source parameter inference \cite{green2020complete,dingo}. \citet{ShenNew} showcases a Bayesian neural network that could produce Bayesian posteriors by repeated sampling with uncertainty of network weights. This method produces accurate posteriors significantly faster than traditional methods, but due to the repeated sampling method, slower than simpler neural network architectures could. \citet{vitamin} use conditional variational autoencoders to accurately predict BBH source parameters, however they have not tested their network on signals embedded in real noise yet. \citet{dingo} applied normalising flows to accurately infer the posteriors of all 15 source parameters, and can produce posteriors that almost perfectly match those produced by LALInference. However, they use frequency domain gravitational wave signals. Frequency domain signals need to accumulate data, which could introduce additional latency. Additionally, in order to achieve this level of accuracy, their network needs to be trained for several weeks, and when predicting must sample the posterior several thousand times which takes several seconds. This may not be fast enough for online parameter inference of GW mergers which will likely be the next target of ML-based inference, to enable rapid EM followups.

In this work, we present a simple system of neural networks which can rapidly predict the chirp mass and mass ratio of a BBH signal in detector noise. BBH signals were used instead of BNS signals, due to the greater length of BNS signals leading to impractically large input layers. These networks are designed for speed and simplicity, meaning they can be rapidly re-trained and are are well suited for integration into online search pipelines \cite{chichi}. Our method produces a mean and standard deviation prediction for each parameter (thus assuming a Gaussian distribution), which can be used to avoid the computational cost of sampling to produce a posterior, with no loss in prediction power over Bayesian prediction methods. By estimating both the chirp mass and mass ratio, it is possible to estimate the two component masses, $m_1$ and $m_2$. We apply this system to the 10 BBH events from the first two gravitational wave observing runs, and report on the network's performance.

\section{Network architecture}

The two neural network prediction systems presented here for chirp mass and mass ratio respectively use the same general architecture, with slightly different parameters and training as outlined in section \ref{results}. Both prediction systems consist of a denoising autoencoder, whose output is passed to a dense multi-branched convolutional neural network which predicts the relevant parameter.

The first component of both systems is a denoising autoencoder \citep{ae}, which cleans the GW time series data from LIGO Hanford and Livingston detectors to remove as much noise as possible. Autoencoders are neural networks which aim to reconstruct their own inputs, and are designed to learn a dimension-efficient representation of the input data. An autoencoder is very effective at denoising signals, recovering the `clean' structured waveform while most contaminating noise is removed through the compression of the data. Through trial and error, we found that at high signal-to-noise ratios (SNRs), SNR $>30$ , the dense estimator network can predict the parameters reasonably well on its own. However, as the SNR drops into the level normally seen in real signals, the noise becomes overwhelmingly high and the network struggles to identify the actual waveform among the noise by itself. The denoising autoencoder assists here by handling the noise removal, outputting a cleaner signal which is fed to the estimator network, greatly improving the estimator's performance. Additional information on the architecture, training and performance of the autoencoder can be found at \citet{ae}.

\begin{figure}
    \centering
    \includegraphics[width=0.7\linewidth]{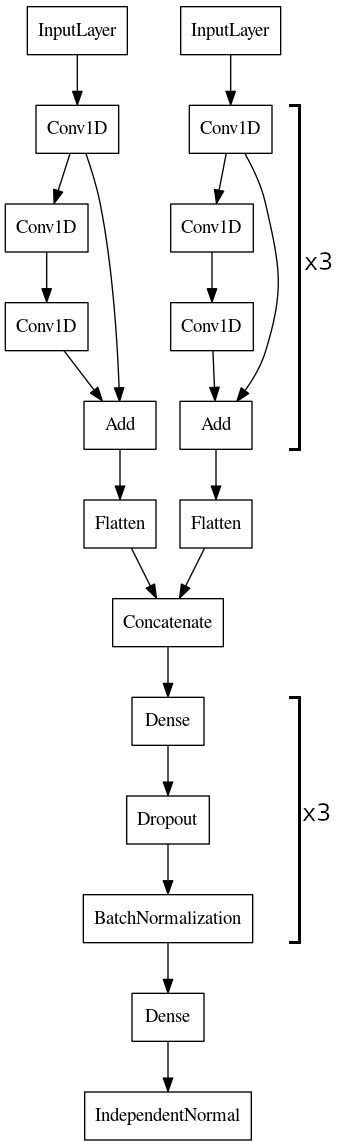}
    \caption{Flow chart representing our CNN design. The input data are the Hanford and Livingston strains that have been denoised by the autoencoder. Additional information on autoencoder architecture can be found in \citet{ae}. Repeated sections of the network have been replaced with a `x3', for simplicity.}
    \label{fig:model}
\end{figure}

 We then apply a second multi-branched dense neural network to predict the parameter of interest from the output of the autoencoder (Fig.~\ref{fig:model}). The network used is a convolutional neural network (CNN), created using the Keras library \cite{keras} with Tensorflow \cite{tensorflow} backend. The network receives denoised GW signals as input, and produces a mean and standard deviation prediction for each signal. First, the detector signals are passed to two separate input branches, which consist of multiple 1D convolution layers and residual blocks. Residual blocks are a type of neural network structure in which the output of a layer `skips' several layers, and is then added to a later output. The advantage of this type of layer structure is that the intervening layers learn the residual of the output and input, which has been shown to improve deep network performance \cite{he2015deep}. These branches are then concatenated into a single branch, which consists of several dense layers. We found that the chirp mass estimator performed best with 256 neurons in the dense layer, while the mass ratio estimator performed best with 64 neurons per layer. Dropout is included in the dense layers to reduce overfitting and to improve the network's performance on unseen signals \cite{hinton2012improving}. After the dense layers, an IndependentNormal Tensorflow layer is used to produce a mean and standard deviation prediction as the network's output. The networks were trained using the ADAM optimiser \cite{kingma2017adam}, and the negative log likelihood loss function was used. Training typically took 1 hour to complete when done using an NVIDIA P100 GPU. 
 
\subsection{Training Set}

Both parameter estimator networks were trained on 500,000 BBH gravitational wave events. These samples were generated using the \texttt{SEOBNRv4} waveform approximant \cite{seobnrv4}, and were injected into Gaussian noise coloured by the detectors' power spectral density (PSD) from the second observing run, O2. These signals were then passed through the autoencoder, with the output of the autoencoder being the training set for the parameter estimator networks. These signals are each a quarter of a second in length and sampled at 2048 Hz, which ensures the waveform covers all detectable BBH signals in our mass range. Each network's training set was uniformly distributed in the relevant parameter (i.e. chirp mass and mass ratio) to avoid the networks becoming biased by the training set. While this is not a realistic astrophysical distribution for real gravitational wave events (high mass ratios are strongly favoured over low mass ratios in detected BBH events \cite{gwtc2}, for example), it ensures the networks are generalised enough to make accurate predictions on future events. Both networks were trained with parameters from Tab.~\ref{tab:parameters}. Note that forcing a uniform distribution in either the chirp mass or mass ratio leads to a non-uniform distribution in both $m_1$ and $m_2$. The coalescence time was not varied in the samples, as it was deemed not relevant for mass estimation.

\begin{table}[h]
    \centering
    \begin{tabular}{cc}
         Parameter& Range  \\
         \hline
         M$_c$ (M$_{\odot}$) &  [6, 70]\\
         q & [0.1, 1]\\
         m$_1$,m$_2$ (M$_{\odot}$)& [7, 100]\\
         distance (Mpc) & [40, 3000]\\
         right ascension & [0, 2$\pi$]\\
         declination & [0, $\pi$]\\
         SNR & [10, 30]\\
         spin & [-1, 0.99]\\
         inclination & [0, $\pi$]\\
         
    \end{tabular}
    \caption{ Parameters used to create the training sets. All non-mass parameters were distributed uniformly, and the chirp mass ($M_c$) and mass ratio (q) were distributed uniformly for their respective networks. The target SNR was achieved by calculating a scaling factor from the optimal matched filtering SNR, then scaling the signal amplitude by this factor~\cite{gebhard}.}
    \label{tab:parameters}
\end{table}

The chirp mass and mass ratio of a gravitational wave event are defined in Equations \ref{cm_equation},\ref{q_equation}. These parameters determine the frequency evolution of the compact binary coalescence (CBC), and are therefore easier to measure from the gravitational wave signal than the individual masses, $m_1$ and $m_2$ (where $m_1 > m_2$) \citep{CM_inspiral}. The component masses can still be recovered by solving these chirp mass and mass ratio equations for $m_1$ and $m_2$.

\begin{equation}
\label{cm_equation}
    M_c = \frac{(m_1m_2)^{\frac{3}{5}}}{(m_1+m_2)^{\frac{1}{5}}}
\end{equation}

\begin{equation}
\label{q_equation}
    q = \frac{m_2}{m_1}
\end{equation}

\section{Prediction Results}
\label{results}

\begin{figure}
\includegraphics[width=\linewidth]{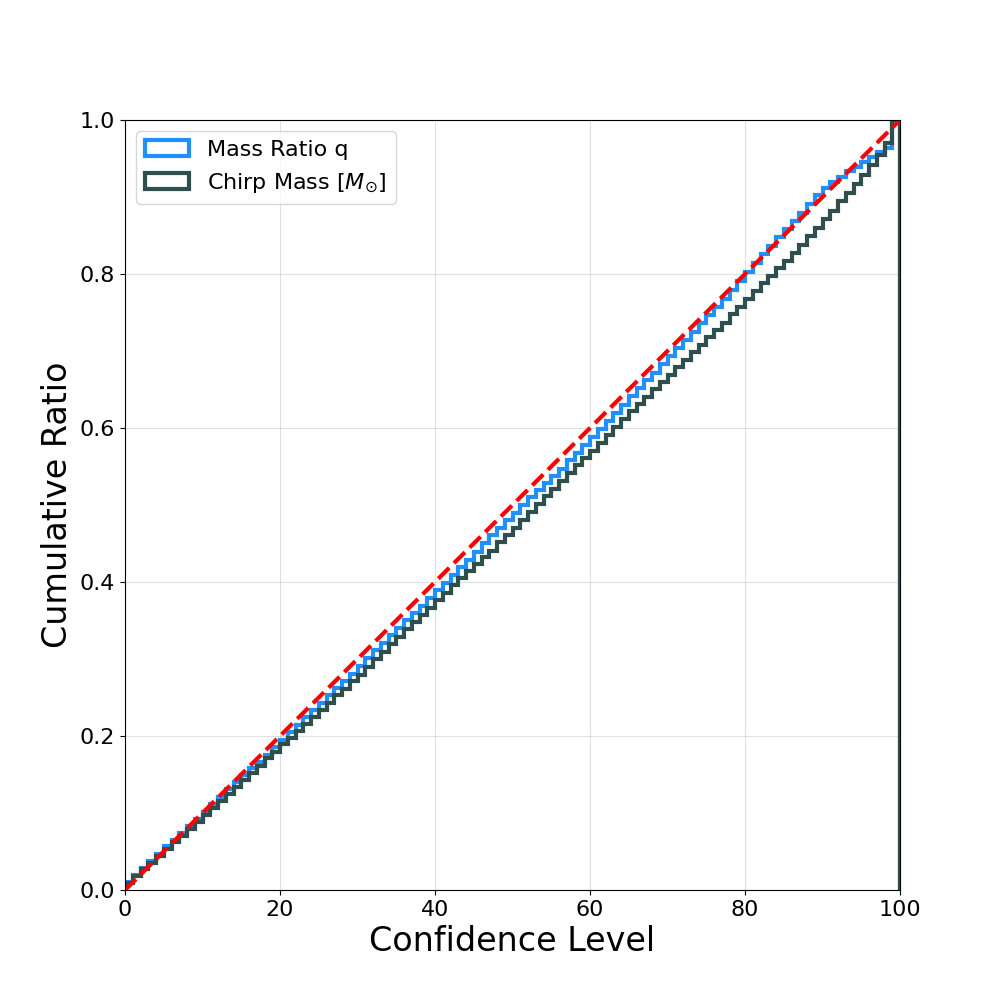}
\caption{\label{fig:pp-plot}p-p plot for the 25,000 predictions made on the Gaussian test set for both parameters. Each curve shows the cumulative distribution of the confidence levels containing the true parameter.}
\end{figure}

The parameter estimator systems were tested on a set of 25,000 simulated signals with SNRs between 10 and 30, which were not part of the training set. The signals were injected into Gaussian noise coloured by the O2 PSD. The network's ability to produce fast and accurate predictions and useful probability distributions were tested. The networks were able to produce a prediction for each sample in $\sim$ 1.4 milliseconds when tested with a NVIDIA P100 GPU and an Intel Gold 6140 CPU, a prediction speed which is faster than that shown by \citet{ShenNew}, \citet{ChuaAlvinJ} and \citet{GreenStephenR2020Gpew}. The networks were shown to produce useful confidence intervals using a probability-probability plot (Fig.~\ref{fig:pp-plot}). This plot was constructed by taking the smallest confidence interval produced by the networks that contained the true value for each sample, and then binning the data and producing a cumulative distribution. This shows that the networks are able to produce useful confidence intervals as every confidence level contained a number of true parameters consistent with the expected count, within 5$\%$. 

\begin{figure}
\includegraphics[width=\linewidth]{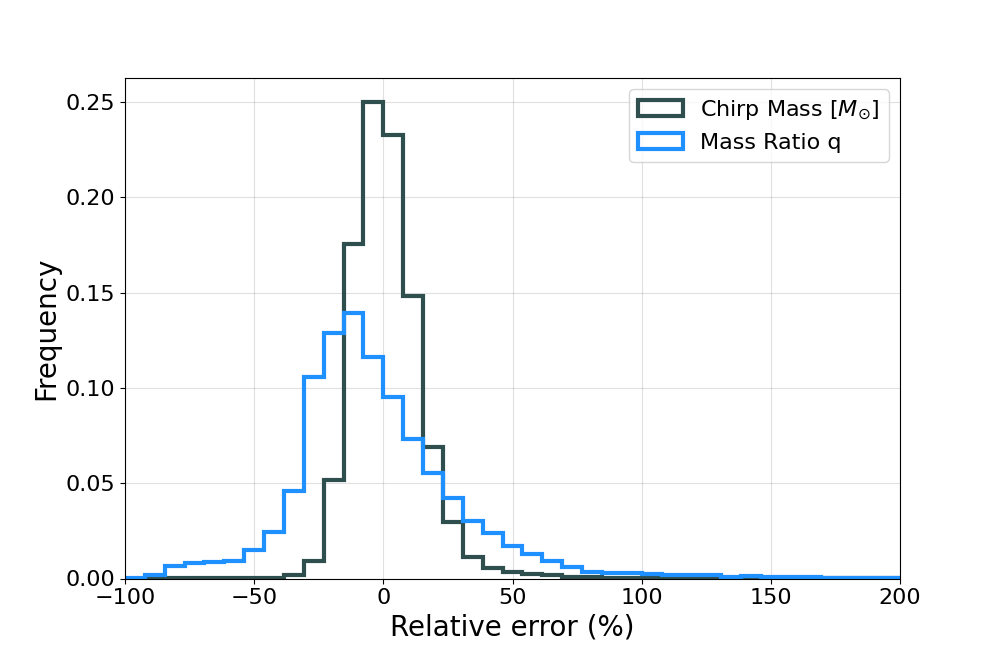}
\caption{\label{fig:error_hists}Error histograms for the two networks. Note that the average error on the mass ratio network is higher, and skewed towards underestimating the mass ratio.}
\end{figure}

The chirp mass estimator had a lower average error than the mass ratio estimator (Fig.~\ref{fig:error_hists}), which was expected as the mass ratio of a signal is not as strongly constrained as its chirp mass \citep{CM_inspiral}. The chirp mass estimator showed no significant bias in overestimating/underestimating, with a 51/49 split. The mass ratio estimator showed a slight bias toward underestimating, however, with a 40/60 split. This can be seen in the error histograms for both networks, as the chirp mass predictions appear symmetrically distributed, while the mass ratio predictions are skewed towards underestimating.

\begin{figure}
    \centering
    \includegraphics[width=\linewidth]{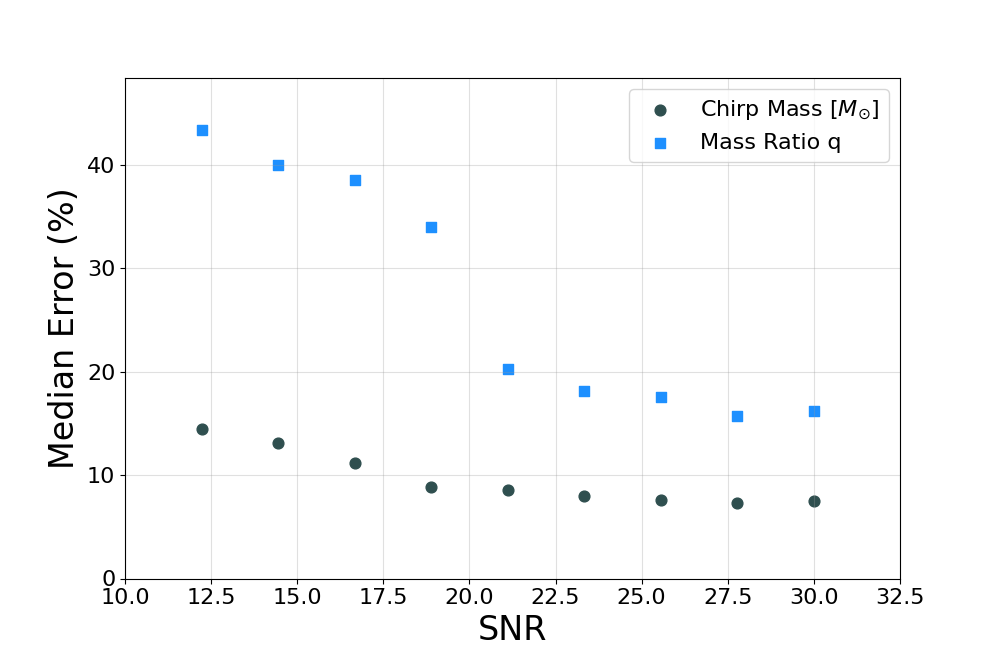}
    \caption{Binned median parameter estimator prediction error versus network SNR on test set.}
    \label{fig:error_snr}
\end{figure}

As expected, the accuracy of both networks improved with increasing SNR (Fig.~\ref{fig:error_snr}). This shows that the networks perform better on higher SNR signals, presumably due to the autoencoder producing more accurate waveforms. Since the SNR of a signal is dependent on the chirp mass (due to the chirp mass affecting the signal strength), this improvement may partially be due to the chirp mass of the signal as seen in Fig~\ref{fig:cm_error}.

\subsection{Estimator network results}
\begin{figure}
\includegraphics[width=\linewidth]{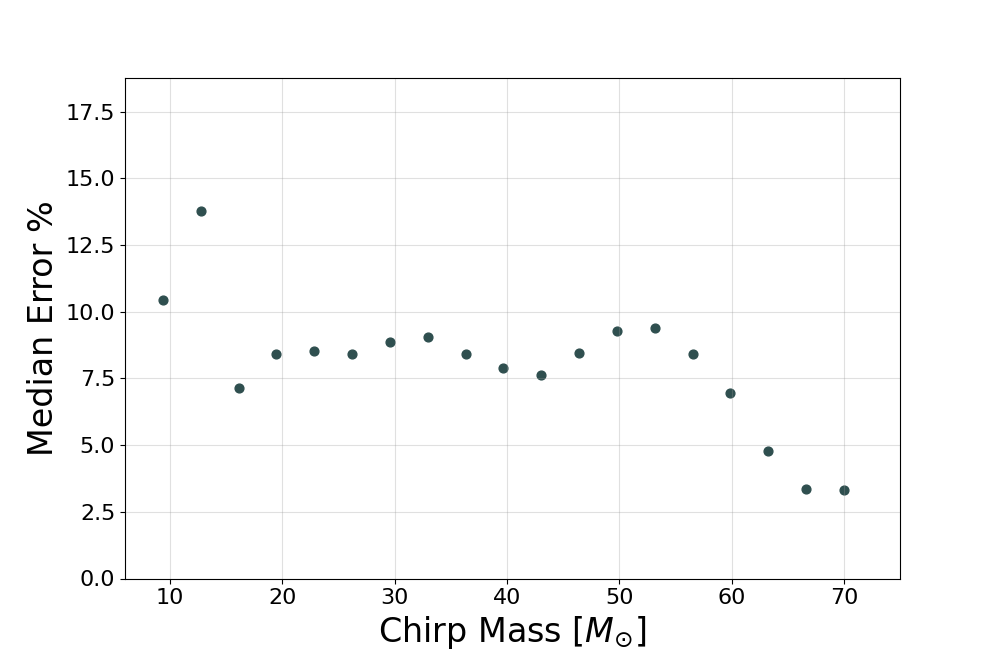}
\caption{\label{fig:cm_error}Binned median chirp mass error for the Gaussian test set.}
\end{figure}

The chirp mass system achieved a mean error of $10.53\%$ on the Gaussian test set, with the mean error decreasing with increasing chirp mass (see Fig.~\ref{fig:cm_error}). The greater prediction error seen at low chirp masses is partially due to the autoencoder, which produces signal representations with poorer overlaps at lower chirp masses \cite{ae}. Even at low SNRs the network was able to produce predictions with relatively low average error of 16\% compared to the mean of 10.53\%.

The network was then tested on the real data from the 10 O1 and O2 BBH events, that had been passed through the denoising autoencoder. The chirp mass estimator produced predictions with a $6.20\%$ average error between the predicted median and LAL simulations. Every 90\% confidence interval produced by the network contained the true value produced by LAL simulations.

The second estimator network was trained to predict the mass ratio q. The architecture of the network remained largely the same, but with one-quarter the amount of neurons in the dense layers than were used in the chirp mass network.

\begin{figure}
\includegraphics[width=\linewidth]{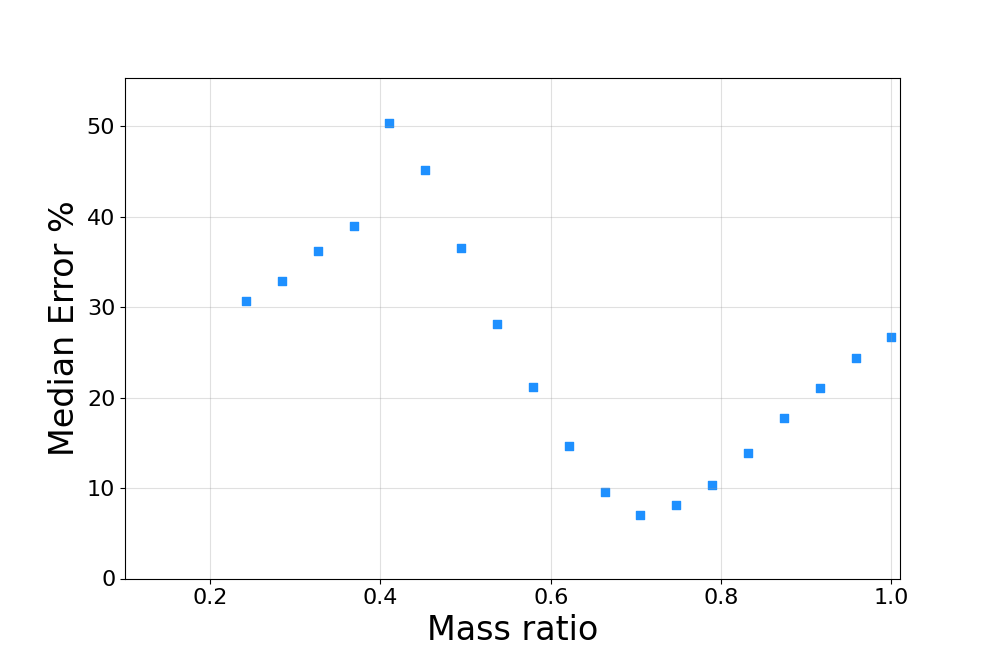}
\caption{\label{fig:q_error}Binned median mass ratio error for the Gaussian test set.}
\end{figure}

The mass ratio system achieved a mean error of $24.26\%$ on the Gaussian test set (Fig.~\ref{fig:q_error}), There was a strong negative correlation with SNR, with prediction accuracy dropping off rapidly below an SNR of 20. We also find that with increasing SNR the network error decreases, as with the chirp mass network (\ref{fig:error_snr}).

The network was then tested on the real data from the 10 O1 and O2 BBH events, that had been passed through the denoising autoencoder. The mass ratio estimator produced predictions with a $8.50\%$ average error between the predicted median and LAL simulations. This error is significantly lower than on the Gaussian test set, due to the 10 real events having predicted mass ratios of $0.56 < q < 0.86$, which corresponds to the lowest error region of the network's predictions (Fig.~\ref{fig:q_error}).

\subsection{Component Masses}

By using the mass ratio and chirp mass predicting networks together it is possible to derive the component masses of a CBC event. This is done by making a prediction for both the mass ratio and chirp mass as described above, and then solving equations \ref{cm_equation} and \ref{q_equation} for the component masses. Repeated sampling is therefore necessary to create a posterior for the component masses, should it be required. The predicted posteriors for the component masses were then compared to the LALInference predictions, with the percent error in the primary mass being $6.73\%$ and the percent error in the secondary mass being $7.87\%$. The median predictions for both masses fell within the LALInference credible intervals for every event, and conversely the predicted 90\% confidence intervals captured the LALInference median predictions for every event. This shows that the networks have learned an accurate representation of the signals that can compete with Bayesian inference.

\begin{figure*}
\includegraphics[width=\linewidth]{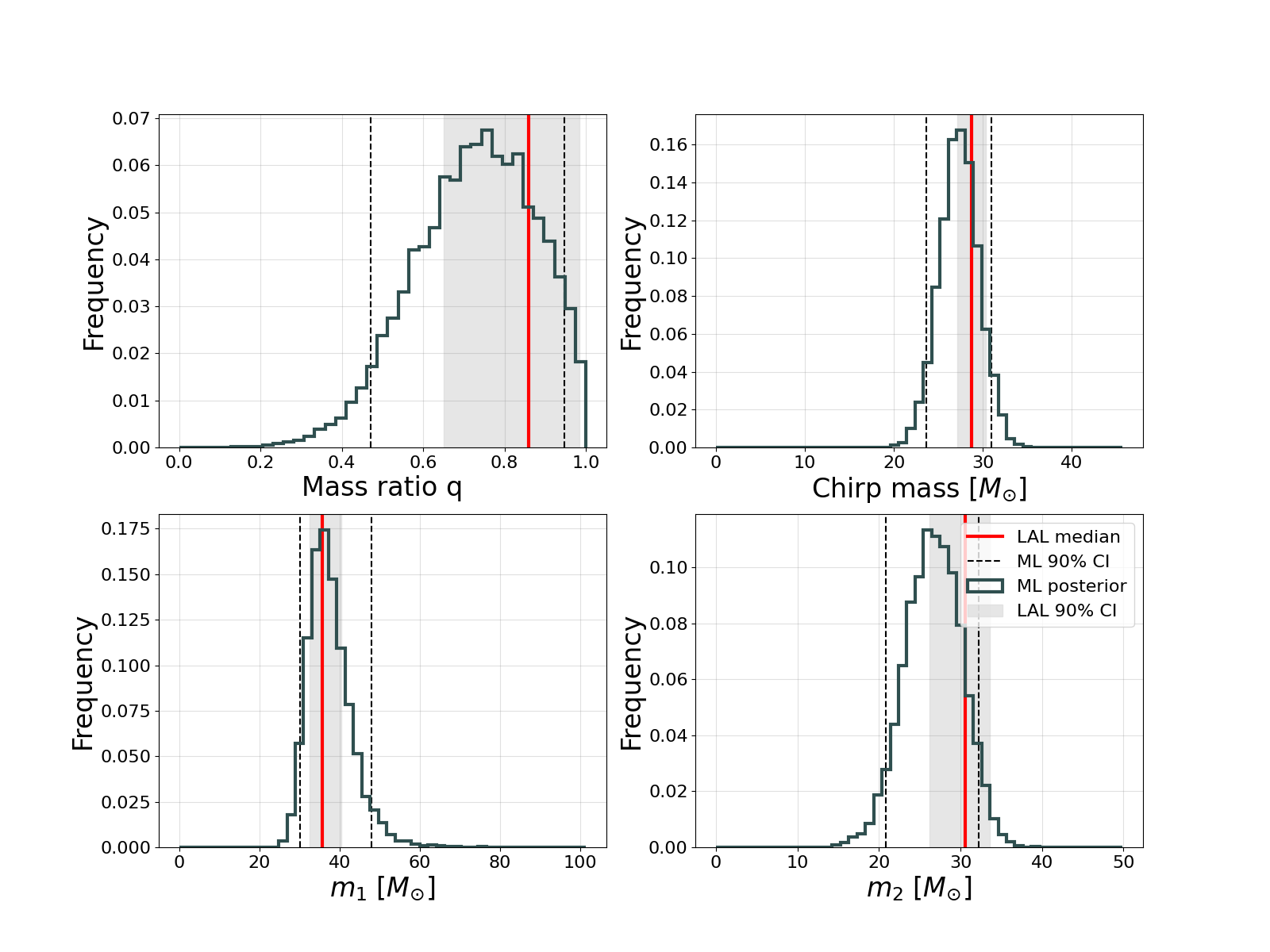}
\caption{\label{fig:150914}Plot of the parameter estimation for GW150914. The solid red vertical lines are the median estimates produced by LAL simulation, and the gray areas are the 90\% credible intervals from the LAL simulations. The vertical dashed lines are the 90\% credible intervals for our predictions. Every plot has a histogram of the network's predictions for that parameter with the component mass predictions being calculated using the chirp mass and mass ratio predictions. These histograms were produced by sampling the chirp mass and mass ratio distributions 10,000 times, as this is necessary to produce the component mass distributions Note that the mass ratio posteriors are truncated at 1 (due to greater mass ratios being unphysical) and 0.1 (due to very small mass ratios leading to extreme component mass predictions). Additional plots of the other O1 and O2 events can be found in the appendix.}
\end{figure*}

A full analysis of the two component mass results are shown in tables \ref{tab:m1} and \ref{tab:m2}. It can be seen in Fig.~\ref{fig:150914} that the predicted credible intervals and median predictions closely match that of the LALInference simulations, with the other event prediction figures available in the appendix.

\begin{table*}[]
\caption{\label{tab:m1}
The results of predicting the component mass $m_1$ on the first 10 BBH events detected by LIGO. The predictions and credible intervals are in units of solar masses ($M_{\odot}$)} 
\begin{ruledtabular}
\begin{tabular}{cccccccc}

 & LAL $m_1$ median & LAL 90\% & $m_1$ deep & Deep learning & & Deep learning &Deep learning \\      
 Signal &prediction     & CI       & learning median    & 90\% CI  & Error \% & median within & cred. interval \\
 & & & & & & LAL cred. bounds? & captures LAL prediction?\\
\hline

GW150914 & 35.6 & [32.6, 40.4] & 36.91 & [30.11, 47.77] & 3.68 & yes & yes\\
GW151012 & 23.3 & [17.8, 37.3] & 21.32 & [16.76, 31.77] & 8.51 & yes & yes\\
GW151226 & 13.7 & [11.5, 22.5] & 15.95 & [11.46, 28.13] & 16.44 & yes & yes\\
GW170104 & 31.0 & [25.4, 38.2] & 27.44 & [22.29, 36.43] & 11.48 & yes & yes\\
GW170608 & 10.9 & [9.2, 16.2] & 11.34 & [7.68, 18.04] & 4.03 & yes & yes\\
GW170729 & 50.6 & [40.4, 67.2] & 50.37 & [38.16, 70.56] & 0.45 & yes & yes\\
GW170809 & 35.2 & [29.2, 43.5] & 32.36 & [26.77, 41.92] & 8.08 & yes & yes\\
GW170814 & 30.7 & [27.7, 36.4] & 31.75 & [26.46, 40.59] & 3.43 & yes & yes\\
GW170818 & 35.5 & [30.8, 43.0] & 33.23 & [26.89, 44.75] & 6.38 & yes & yes\\
GW170823 & 39.6 & [33.0, 49.6] & 37.71 & [30.31, 52.24] & 4.77 & yes & yes\\
 & & & & & 6.73 & 10/10 & 10/10\\

\end{tabular}
\end{ruledtabular}
\end{table*}

\begin{table*}[]
\caption{\label{tab:m2}
The results of predicting the component mass $m_2$ on the first 10 BBH events detected by LIGO. The predictions and credible intervals are in units of solar masses ($M_{\odot}$)}
\begin{ruledtabular}
\begin{tabular}{cccccccc}

 & LAL $m_2$ median & LAL 90\% & $m_2$ deep & Deep learning &  & Deep learning &Deep learning \\      
 Signal &prediction     & CI       & learning median    & 90\% CI  & Error \% & median within & cred. interval \\
 & & & & & & LAL cred. bounds? & captures LAL prediction?\\
\hline

GW150914 & 30.6 & [26.2, 33.6] & 26.68 & [21.06, 32.29] & 12.82 & yes & yes\\
GW151012 & 13.6 & [8.8, 17.7] & 13.57 & [9.47, 17.04] & 0.19 & yes & yes\\
GW151226 & 7.7 & [5.1, 9.9] & 8.55 & [5.32, 11.52] & 11.04 & yes & yes\\
GW170104 & 20.1 & [15.6, 25.0] & 19.15 & [14.73, 23.30] & 4.70 & yes & yes\\
GW170608 & 7.6 & [5.5, 8.9] & 6.69 & [4.18, 9.33] & 12.01 & yes & yes\\
GW170729 & 34.3 & [24.2, 43.4] & 34.54 & [24.67, 44.57] & 0.70 & yes & yes\\
GW170809 & 23.8 & [18.7, 29.0] & 23.22 & [18.19, 27.84] & 2.46 & yes & yes\\
GW170814 & 25.3 & [21.2, 28.2] & 23.45 & [18.66, 27.97] & 7.32 & yes & yes\\
GW170818 & 26.8 & [21.6, 31.1] & 22.93 & [17.50, 28.07] & 14.43 & yes & yes\\
GW170823 & 29.4 & [22.3, 35.7] & 25.58 & [19.15, 31.37] & 12.99 & yes & yes\\
 & & & & & 7.87 & 10/10 & 10/10\\

\end{tabular}
\end{ruledtabular}
\end{table*}

\section{Discussion}

While our networks do not reach the accuracy achieved by other neural network BBH parameter estimators \citep{dingo}, their utility lies in their speed, in both training and prediction. Our estimator networks are able to produce Gaussian posterior predictions for the chirp mass and mass ratio of a CBC within 1.4 milliseconds, and can be trained in an hour on 500,000 GW samples. The component mass errors for these predictions relative to LALInference are within 10\% when tested on real events, and these features combined make our networks ideal to implement in an online search pipeline such as SPIIR \cite{chichi}. This would be especially useful for early warning \cite{magee}, where the rapid classification of GW signals is necessary. Early warning in gravitational wave search pipelines would be of interest for transient astronomers searching for electromagnetic counterparts \citep{clancy}.

The simulated signals used in this work were coloured with the PSD of the detectors during the O2 science run. To improve our results we can train the network on signals coloured by O3 detector noise, and to also test on the BBH signals observed during O3. The three input variant including input from the Virgo detector should also be tested on the dataset coloured with the O3 PSD. It is possible that the inclusion of Virgo may improve the model's performance and bring it even closer to LALInference's level of accuracy.

In this work the neural networks were only trained and tested on BBH systems, however, the LSC has also detected CBCs of systems containing neutron stars (e.g. GW170817). These signals typically exist in the LSC detector sensitivity band for as long as 500 seconds, and thus are much more difficult for a neural network to analyse. However, the neural networks may be able to predict sufficiently small chirp masses on systems containing neutron stars to infer the existence of a neutron star in a signal with minimal modification.

 The denoising autoencoder model produces a single point estimate of a pure GW waveform from noisy strain data. However, there is no unique partition of the data into signal and noise. The CNN is therefore unable to learn the inherent stochasticity in the data distribution, which may explain why our predicted mass posteriors do not exactly match those of LALInference, which models the data distribution as stationary and Gaussian. To improve this, the models could be joined together as a single model, with an additional `uncertainty' parameter being passed to the parameter estimator along with the extracted waveform. This may lead to posteriors which more closely match the LALInference posteriors, as the parameter estimation network would be better equipped to learn the appropriate uncertainty in the relevant parameter.

\section{Conclusion}
This work presented a deep learning system composed of a denoising autoencoder and an estimator that predicts the posterior distribution of the chirp mass of a binary black hole system given the gravitational wave signal emitted during coalescence. The system was trained and tested on binary black hole coalescence gravitational wave signals immersed in Gaussian noise coloured by the PSD of the LVC interferometers during the O2 science run. The system has an average error of $10.53\%$ for chirp mass and $24.26\%$ for the mass ratio across the test set of signals of SNRs between 10 and 30, but is several orders of magnitudes faster than LALInference, taking $\sim$ 1 millisecond to make a prediction on each event. This work shows the promise of deep learning applied to gravitational wave parameter estimation by approaching the accuracy of LALInference despite being much lower latency. This shows that deep learning can take on the role of a rapid first estimation for parameters while the costly LALInference estimation can be performed offline for a more accurate estimate. 

This work could be extended to estimating the chirp mass and mass ratio of neutron star binary systems with minor changes to the input dimensions. The system design presented here should also be trained and tested using real detector noise from the O3 science run. If the system also performs well on real detector noise or can be extended to do so, it will be ready to be implemented in the LVC parameter estimation pipeline as a rapid initial parameter estimator.

\section{Acknowledgements}

AM and DJ utilized the OzSTAR national facility at Swinburne University of Technology. The OzSTAR program receives funding in part from the Astronomy National Collaborative Research Infrastructure Strategy (NCRIS) allocation provided by the Australian Government. AM and DJ acknowledge support from  Australian Research  Council  Centre  of  Excellence for  Gravitational  Wave  Discovery (OzGrav).

We thank Le\"{i}la Haegel and Stephen Green for the helpful comments on the paper.

This research has made use of data, software and/or web tools obtained from the Gravitational Wave Open Science Center (\url{https://www.gw-openscience.org/}), a service of LIGO Laboratory, the LIGO Scientific Collaboration and the Virgo Collaboration. LIGO Laboratory and Advanced LIGO are funded by the United States National Science Foundation (NSF) as well as the Science and Technology Facilities Council (STFC) of the United Kingdom, the Max-Planck-Society (MPS), and the State of Niedersachsen/Germany for support of the construction of Advanced LIGO and construction and operation of the GEO600 detector. Virgo is funded, through the European Gravitational Observatory (EGO), by the French Centre National de Recherche Scientifique (CNRS), the Italian Istituto Nazionale di Fisica Nucleare (INFN) and the Dutch Nikhef, with contributions by institutions from Belgium, Germany, Greece, Hungary, Ireland, Japan, Monaco, Poland, Portugal, Spain. This material is based upon work supported by NSF's LIGO Laboratory which is a major facility fully funded by the National Science Foundation.

CC, LW and FP acknowledge funding and support  from  Australian Research  Council  Centre  of  Excellence for  Gravitational  Wave  Discovery (OzGrav)  under grant CE170100004.

\section{Appendices}

\begin{figure*}
\includegraphics[width=\linewidth]{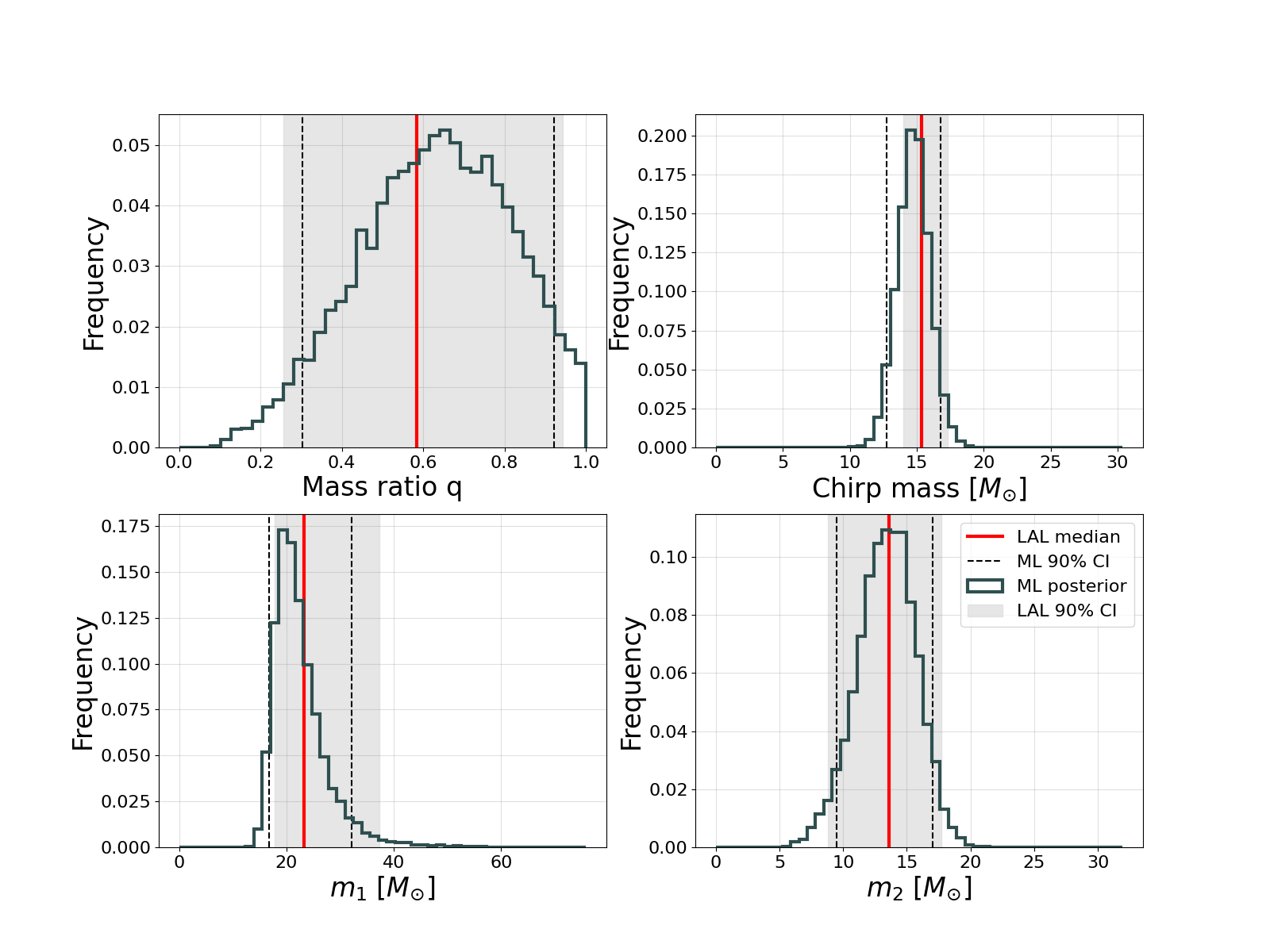}
\caption{\label{fig:151012}Plot of the parameter estimation for GW151012.}
\end{figure*}

\begin{figure*}
\includegraphics[width=\linewidth]{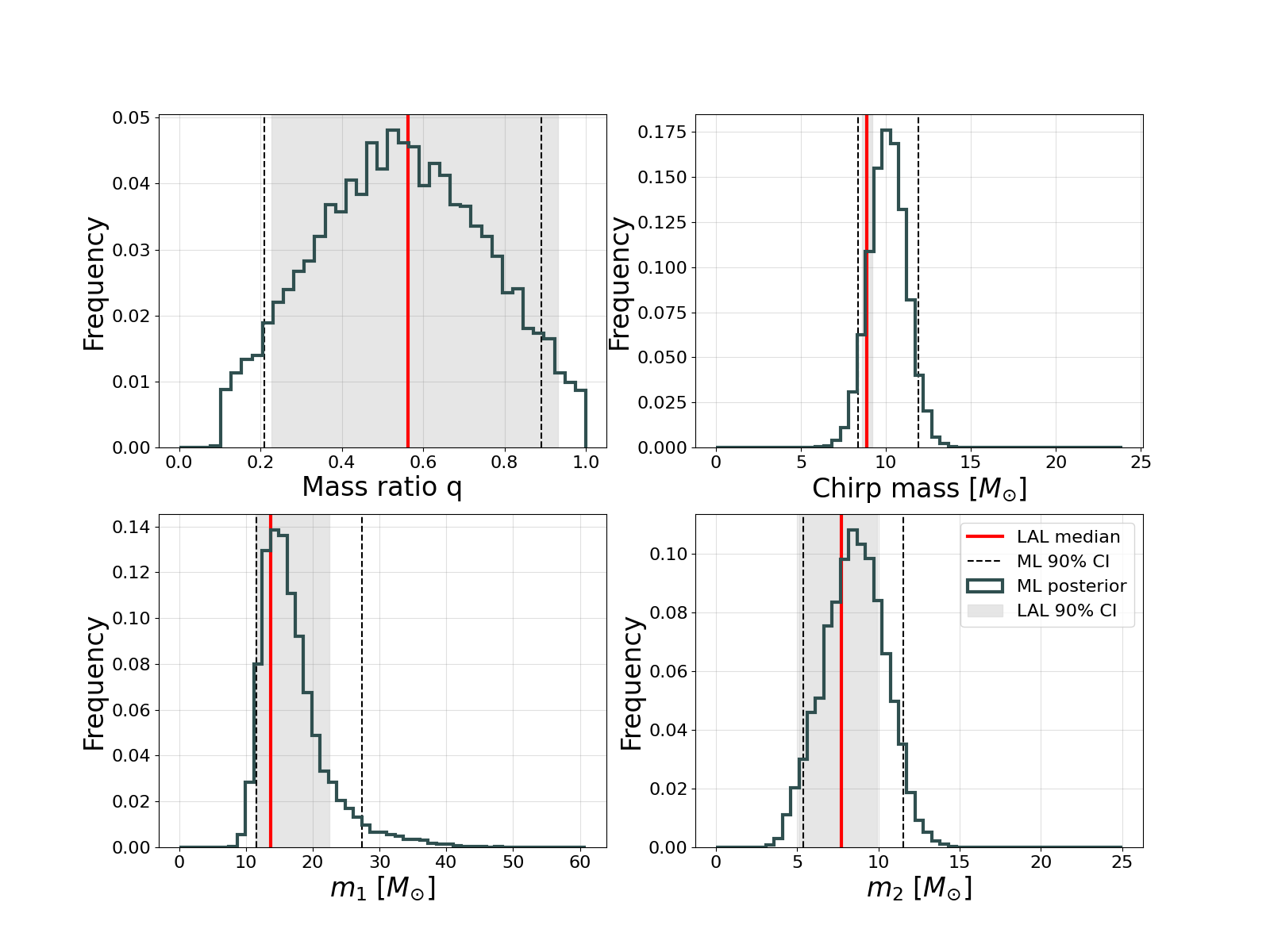}
\caption{\label{fig:151226}Plot of the parameter estimation for GW151226.}
\end{figure*}

\begin{figure*}
\includegraphics[width=\linewidth]{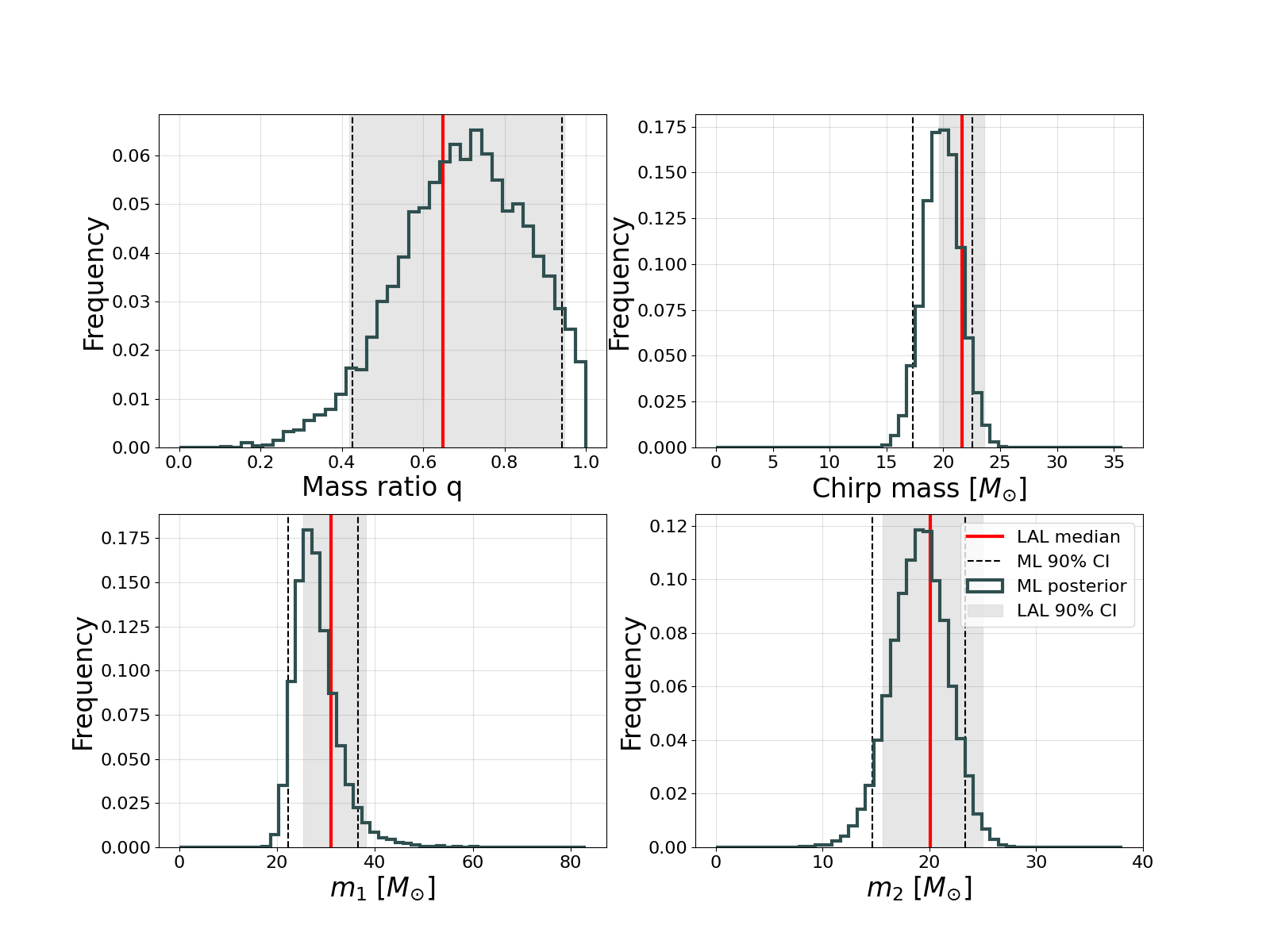}
\caption{\label{fig:170104}Plot of the parameter estimation for GW170104.}
\end{figure*}

\begin{figure*}
\includegraphics[width=\linewidth]{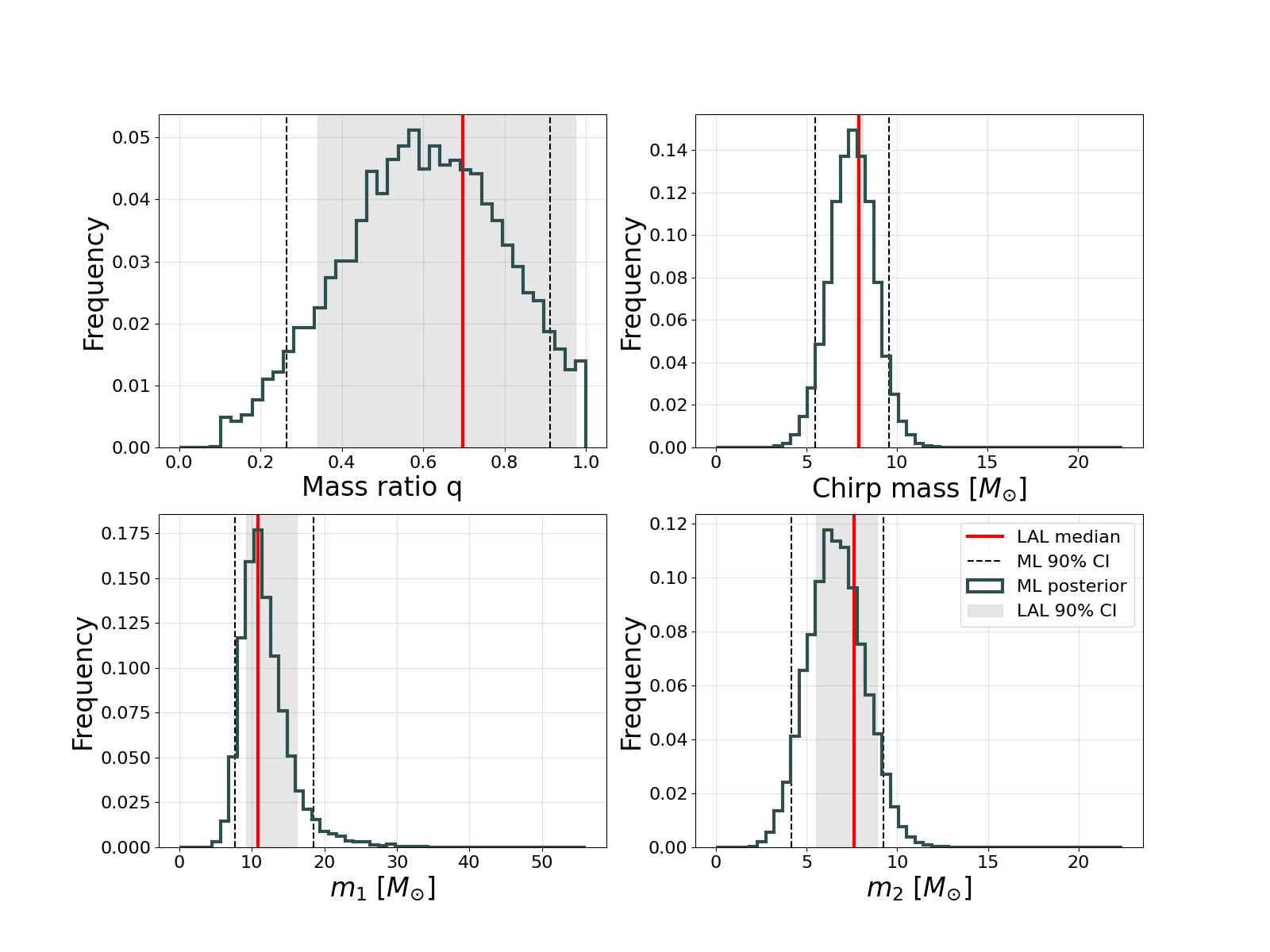}
\caption{\label{fig:170608}Plot of the parameter estimation for GW170608.}
\end{figure*}

\begin{figure*}
\includegraphics[width=\linewidth]{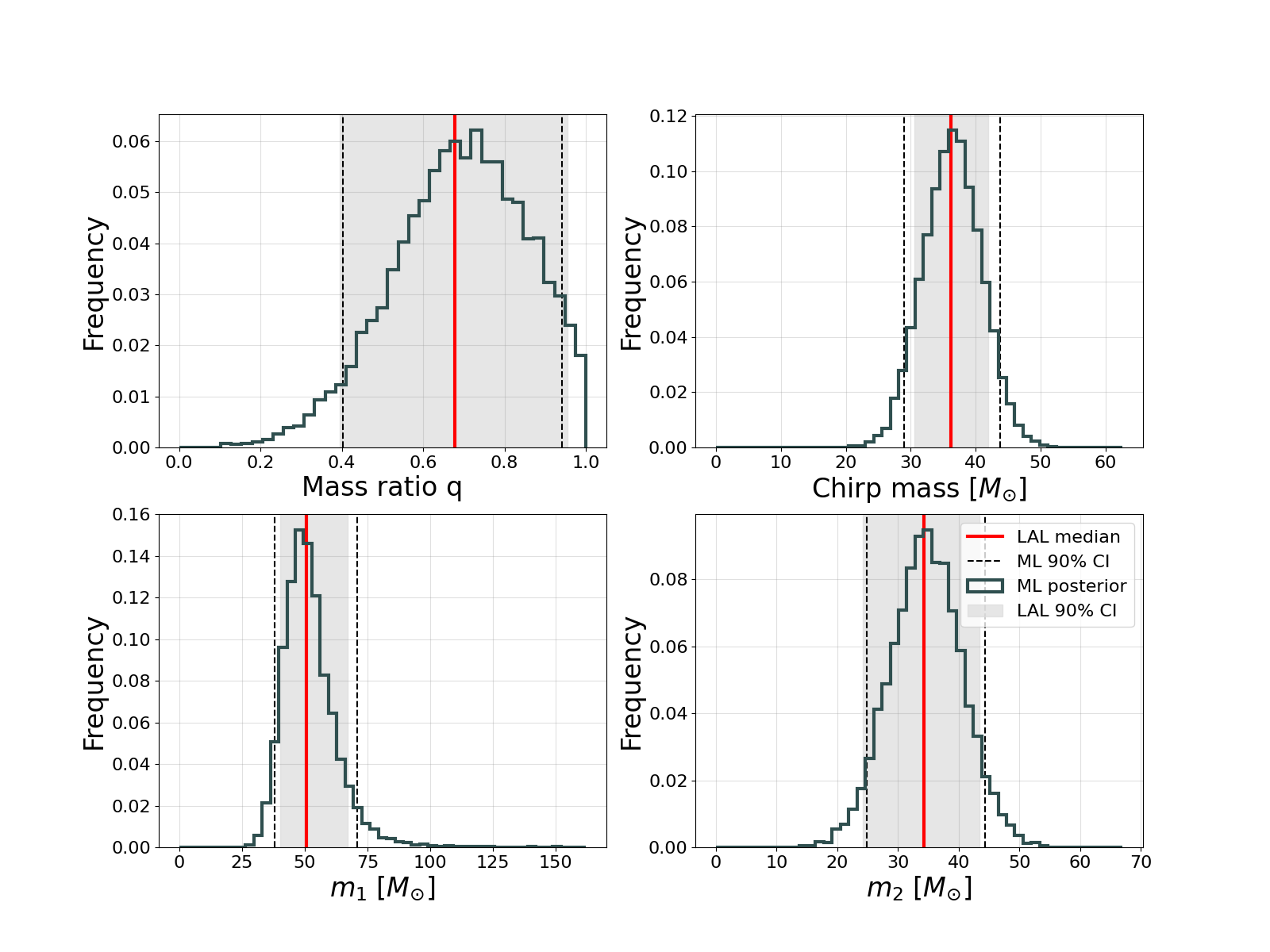}
\caption{\label{fig:170729}Plot of the parameter estimation for GW170729.}
\end{figure*}

\begin{figure*}
\includegraphics[width=\linewidth]{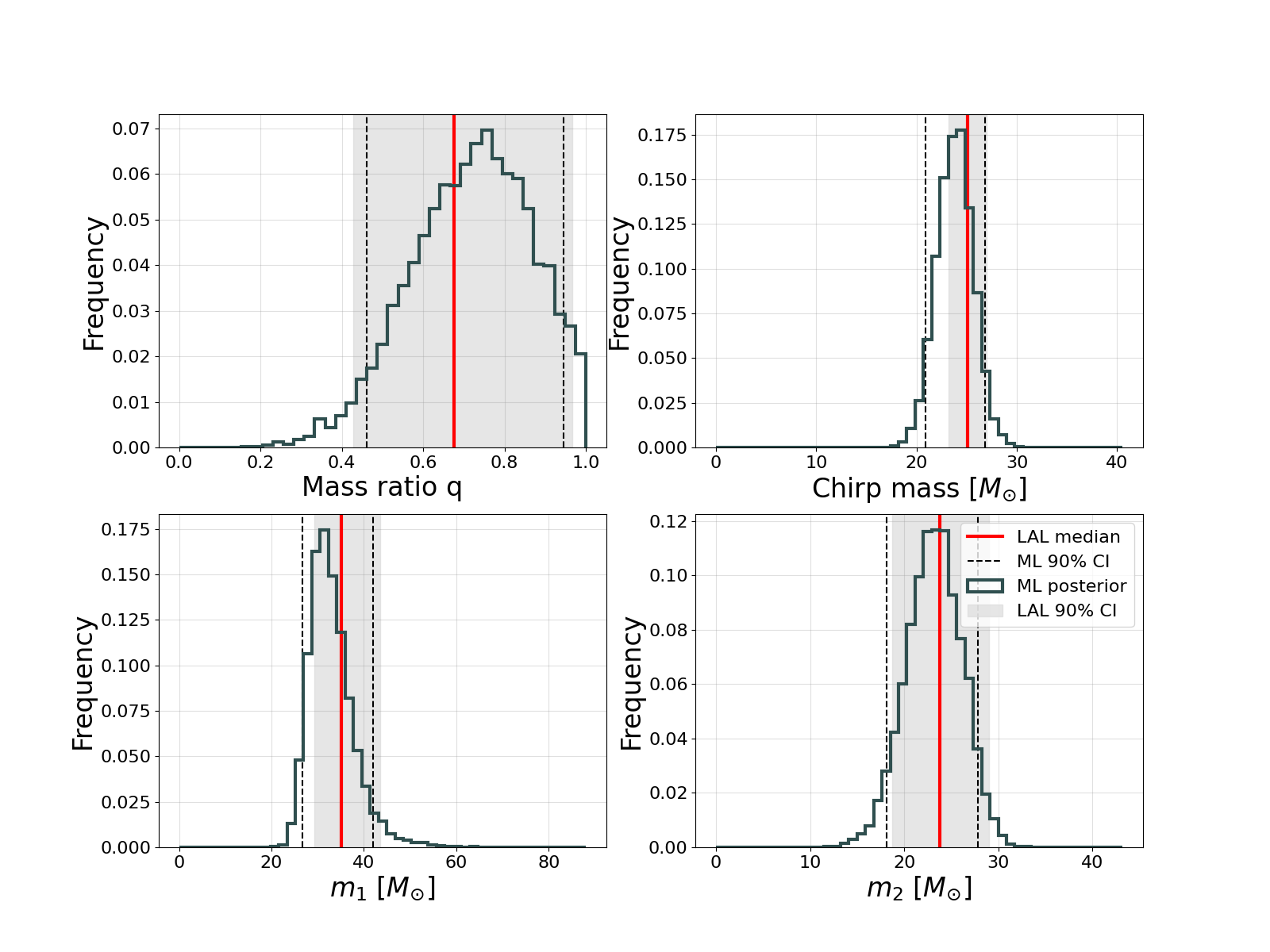}
\caption{\label{fig:170809}Plot of the parameter estimation for GW170809.}
\end{figure*}

\begin{figure*}
\includegraphics[width=\linewidth]{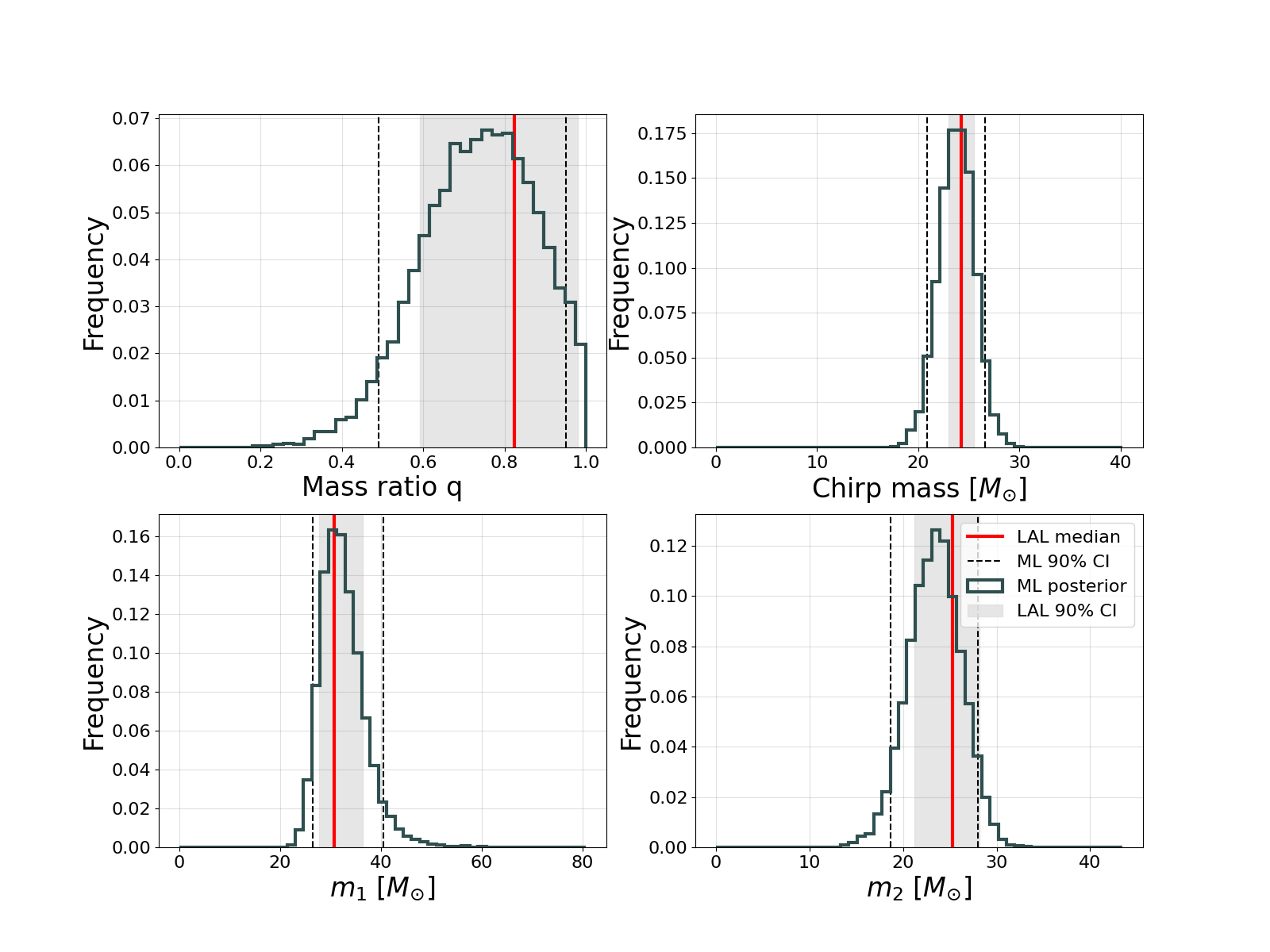}
\caption{\label{fig:170814}Plot of the parameter estimation for GW170814.}
\end{figure*}

\begin{figure*}
\includegraphics[width=\linewidth]{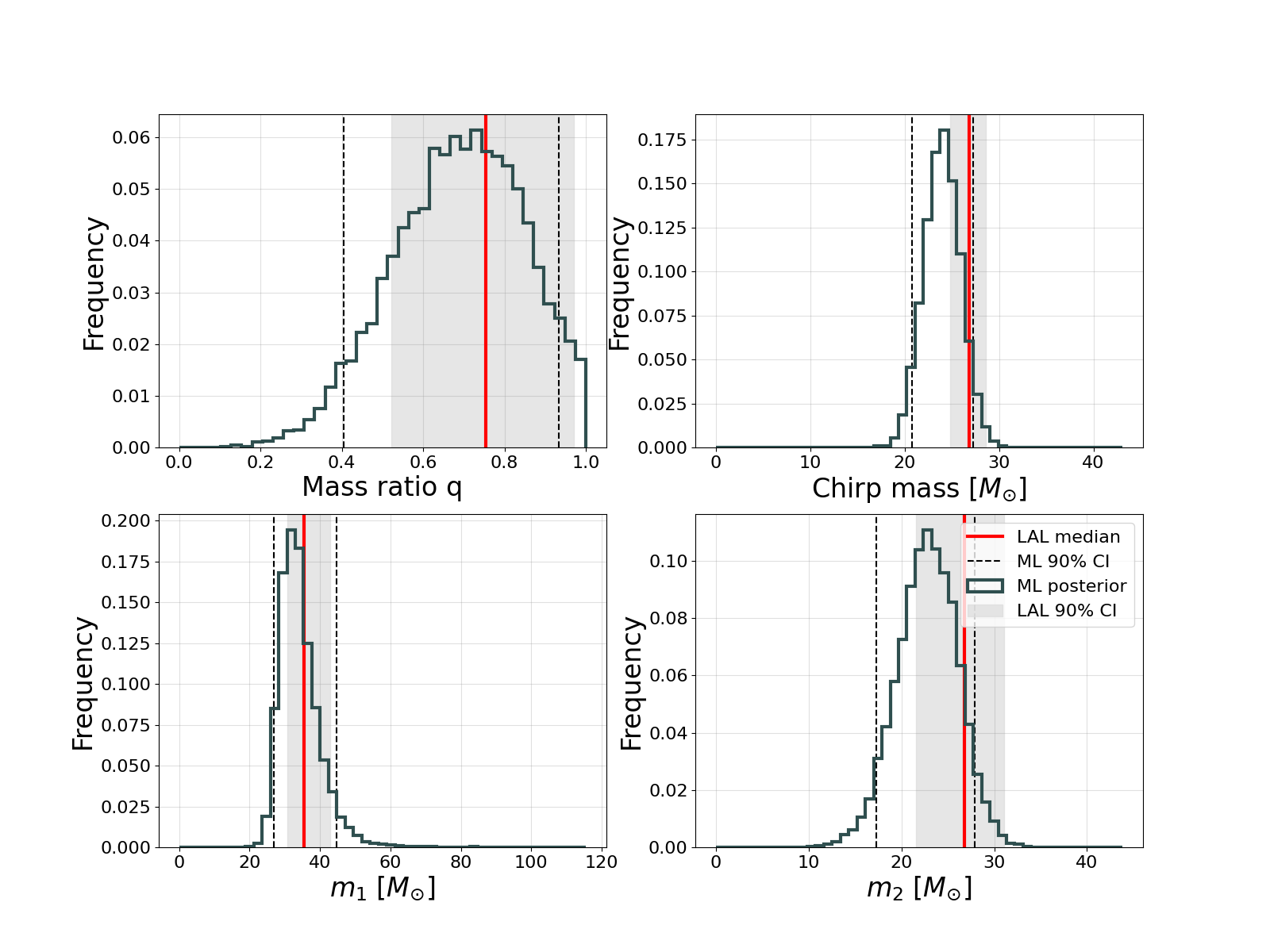}
\caption{\label{fig:170818}Plot of the parameter estimation for GW170818.}
\end{figure*}

\begin{figure*}
\includegraphics[width=\linewidth]{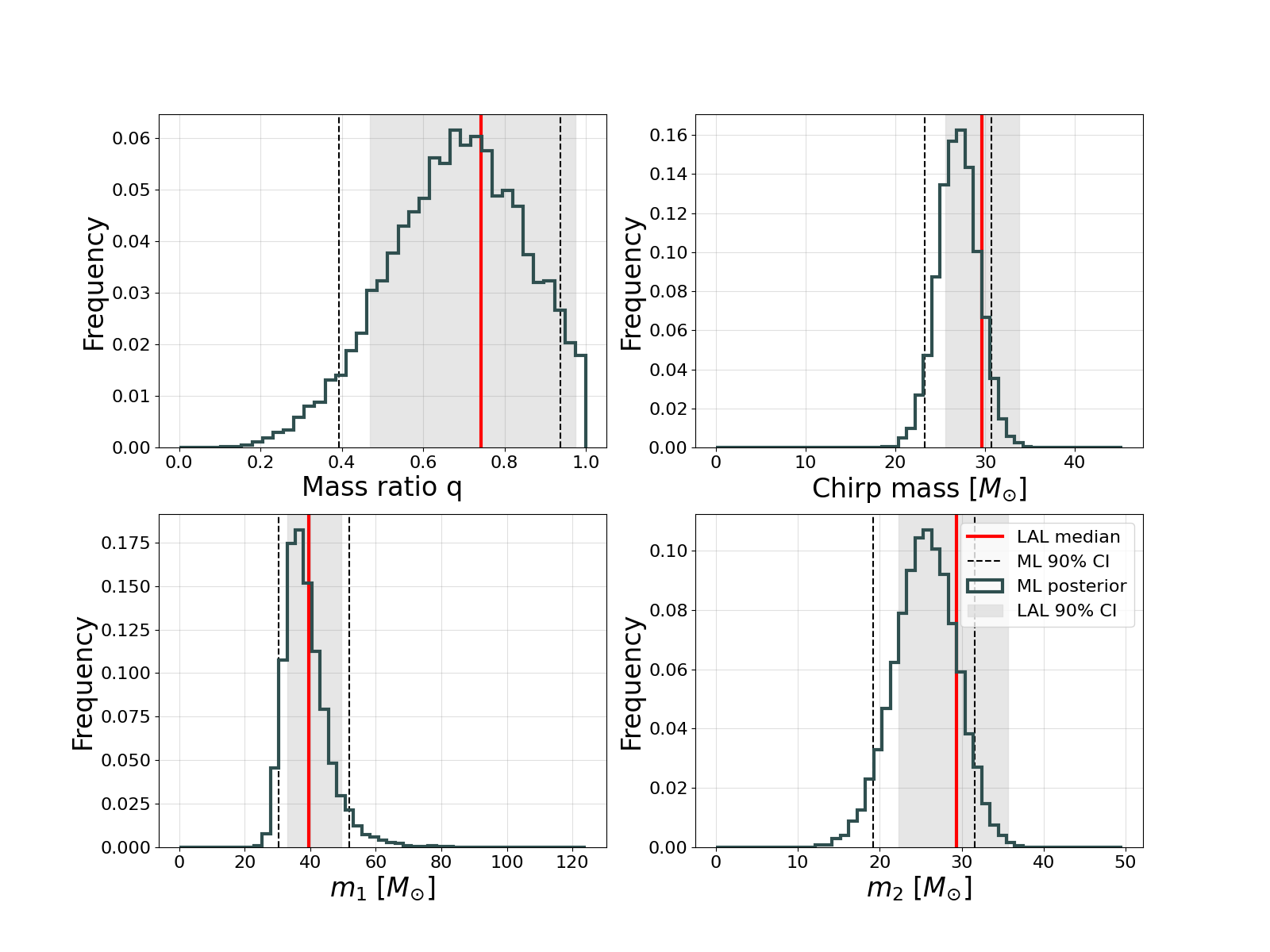}
\caption{\label{fig:170823}Plot of the parameter estimation for GW170823.}
\end{figure*}

\bibliography{bibliography}

\end{document}